\documentclass[english,twocolumn]{revtex4}
\usepackage{babel}
\usepackage{float}
\usepackage{graphicx}
\usepackage{epstopdf}
\usepackage{amsmath}
\usepackage{amssymb}
\usepackage{esint}
\usepackage{hyperref}
\usepackage{natbib}
\usepackage{color}
\usepackage{setspace}

\begin{document}

\title{Fast nanoscale addressability of nitrogen-vacancy spins via coupling to a dynamic ferromagnetic vortex}

\author{M. S. Wolf$^1$ , R. Badea$^1$,  J. Berezovsky$^1$}

\affiliation{$^1$Department of Physics, Case Western Reserve University, Cleveland,
	Ohio 44106. Correspondence and requests should be addressed to J. Berezovsky (email: jab298@case.edu)}
	
\begin{abstract}
The core of a ferromagnetic vortex domain creates a strong, localized magnetic field which can be manipulated on nanosecond timescales, providing a platform for addressing and controlling individual nitrogen-vacancy center spins in diamond at room temperature, with nanometer-scale resolution. First, we show that the ferromagnetic vortex can be driven into proximity with a nitrogen-vacancy defect using small applied magnetic fields, inducing significant nitrogen-vacancy spin splitting.  Second, we find that the magnetic field gradient produced by the vortex is sufficient to address spins separated by nanometer length scales. By applying a microwave-frequency magnetic field, we drive both the vortex and the nitrogen-vacancy spins, resulting in enhanced coherent rotation of the spin state. Finally we demonstrate that by driving the vortex on fast timescales, sequential addressing and coherent manipulation of spins is possible on $\sim100$~ns timescales.
\end{abstract}

\maketitle
Future spintronic devices will require fast, strong, localized magnetic fields (or effective magnetic fields) in an integrated platform\cite{Dietl2008}. Ferromagnetic (FM) vortices provide a route towards this requirement, in particular, when coupled to nitrogen-vacancy (NV) center spins in diamond. NVs are an increasingly attractive candidate for applications in both spin-based sensing\cite{Maze2008,Balasubramanian2008,Degen2008} and quantum information processing\cite{Dutt2007,Wrachtrup2006} due to their long coherence times at room temperature\cite{Balasubramanian2009a,Childress2006,Jelezko2004}, and nanoscale size. But to take advantage of the NV's small size, individual spins must be addressable on nanometer length scales, for individual manipulation and read-out of the spin state, and to control coupling between proximal spins.

Previous work has studied the possibility of using a magnetic field gradient for spin addressability, though without the necessary combination of nanometer-scale resolution, fast control, and potential scalability. By using the static fringe field from a uniformly magnetized micromagnet, spin addressability was demonstrated in coupled gate-defined quantum dots with comparatively large size scale\cite{Pioro-Ladriere2008}. Confined spin-wave modes in such micromagnets have been shown to locally couple to NV spins\cite{VanderSar2015}. But for nanoscale NV addressability, larger magnetic field gradients are required than are typically generated by a uniformly magnetized micromagnet. One approach is to use a scanning nanomagnet (i.e. a magnetic force microscopy tip) to provide the necessary gradient\cite{Grinolds2011,Myers2014}. The inverse process, scanning NV magnetometry, in which an NV center is placed on a scanning probe, has proven to be a powerful technique for revealing the fringe fields from magnetic structures, including vortex domains\cite{Taylor2008,Rondin2013a,Tetienne2013}. These scanning probe approaches do not permit fast control. Here, we show that a FM vortex can provide a large local magnetic field and magnetic field gradient for addressing spins, while also being able to rapidly drive the vortex position for coherent control of the NV spin. 

\begin{figure}[tbp]
	\centering
	\includegraphics[width=0.48\textwidth]{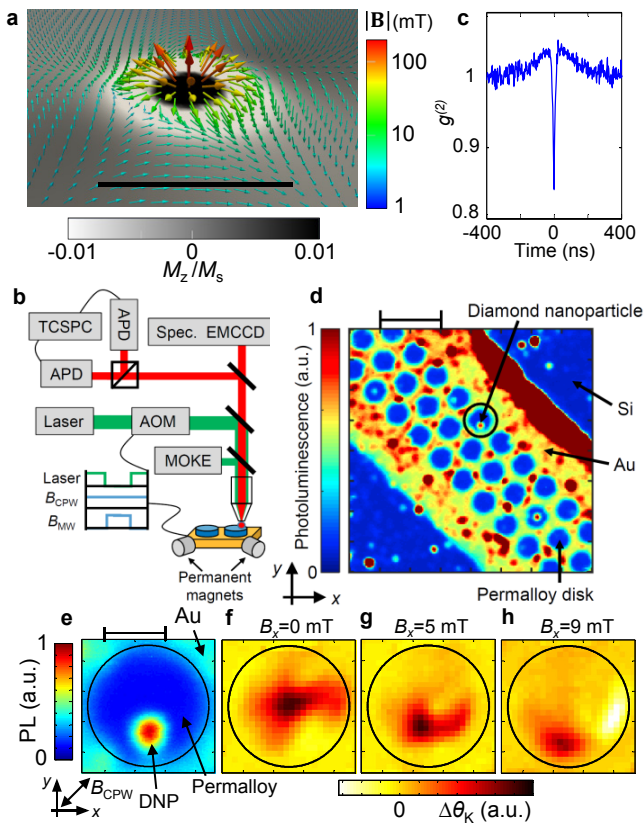}
	\caption{\textbf{Characterization of the NV/vortex system. a},  Micromagnetic simulation near the vortex core with in-plane applied field $B=3$~mT showing normalized out-of-plane magnetization $M_{z}/M_{\mathrm{s}}$ (gray scale), and the magnetic field 20~nm above the surface (arrows, logarithmic scale). Scale bar, 100 nm. $\textbf{b}$, Schematic of the setup. A continuous-wave 532 nm laser is sent through an acousto-optical modulator (AOM) for pulsing. The reflected laser from the sample is sent to a magneto-optical Kerr effect microscopy (MOKE) setup for imaging the vortex core.  The PL is directed to either an electron-multiplied CCD (EMCCD) camera for imaging or sent to a pair of avalanche photodiodes (APDs) connected to a time-correlated single photon counter (TCSPC).  $\textbf{c}$, A $g^{(2)}$ measurement of the nanoparticle indicating multiple NV centers. $\textbf{d}$, PL scan of the sample. Scale bar, 5 $\mu$m.  $\textbf{e}$,PL map of the NV/vortex system studied here. The circle indicates the edge of the Permalloy disk, and the double arrow indicates the direction of $B_{\mathrm{CPW}}$. Scale bar, 1 $\mu$m.  $\textbf{f-g}$, Differential magneto-optical Kerr effect measurements showing the vortex core positions at different applied fields $B_{x}$ = 0, 5, and 9 mT, respectively.}
	\label{setup}
\end{figure} 

\section*{\textbf{Results}}

\textbf{Characterizing the NV/vortex system.}  The core of a ferromagnetic vortex provides a strong, local, controllable magnetic field for spin addressability and control. The ground state magnetization of a FM microdisk is a single vortex domain with in-plane magnetization circulating around a central core of magnetization normal to the plane\cite{Shinjo2000,Cowburn1999,Wachowiak2002}. A simulation of vortex magnetization near the core, and the resulting fringe field, is shown in Fig.~\ref{setup}a. The gray scale represents the normalized out-of-plane magnetization $M_{z}/M_{\mathrm{s}}$ where $M_{\mathrm{s}}$ is the saturation magnetization. The scale is oversaturated (completely black) at the central vortex core, where $M_{z} \approx M_{\mathrm{s}}$, in order to show more detail away from the core. The arrows show the magnetic field in a plane 20~nm above the disk surface, with the arrow length and color showing the field strength on a logarithmic scale. The field is strongest directly above the vortex core, and then falls off approximately as  $r^{-3}$ with small fields persisting beyond the core due to long-range deformation of the vortex domain. 

To demonstrate the vortex-enabled spatial addressability we need two or more NVs with nm-scale separation. Diamond nanoparticles (DNPs) provide a straightforward way to achieve this. We have used DNPs with mean diameter of 25 nm, containing typically one to several NVs. NVs are detected optically in a scanning microscopy setup illustrated in Fig.~\ref{setup}b. Figure~\ref{setup}c shows the two-photon correlation $g^{(2)}$ for the particular DNP we study here. The central dip in $g^{(2)}<5/6$, indicating that there are roughly 5 or fewer NVs in this DNP.  Because the metal substrate also generates some background counts, it is probable that the DNP contains three or four NVs.

NVs and FM vortices are coupled by overcoating a set of Permalloy disks with DNPs. To apply both a MW magnetic field with amplitude $B_{\mathrm{MW}}$ and a rapidly-tunable static magnetic field $B_{\mathrm{CPW}}$, the disks are fabricated atop a gold coplanar waveguide (CPW) as shown in the photoluminescence (PL) scan in Fig.~\ref{setup}d.  Diagonally across the scan is a portion (center) of the CPW. The CPW center conductor constricts to a 10 $\mu$m width for 100 $\mu$m at the middle where the Permalloy (Ni$_{0.81}$Fe$_{0.19}$) disks are seen as the array of blue circles. The gold, Permalloy, and silicon are distinguishable in the PL scan due to differing background PL counts on these materials. DNPs are dispersed over the entire surface. DNPs generate diffraction limited spots across the image.  Their PL intensity is dependent on the number of NVs within the DNP having $\approx$ 1 to several NVs per particle.  The specific DNP/vortex system studied is circled in Fig.~\ref{setup}d and a zoomed in PL scan is shown in Fig.~\ref{setup}e.  The black circle indicates the edge of a 2-$\mu$m-diameter, 40-nm-thick Permalloy disk. The DNP appears as the bright spot in the lower half of the FM disk. 

The vortex core position $\mathbf{r}_{\mathrm{v}}$ is controlled by the application of an in-plane magnetic field $\mathbf{B}$. To lowest order, $|\mathbf{r}_{\mathrm{v}}| = \chi_0 |\mathbf{B}|$, with $\mathbf{r}_{\mathrm{v}} \perp \mathbf{B}$\cite{Guslienko2001,Burgess2014}. Differential scanning magneto-optical Kerr effect (MOKE) microscopy is used to image the vortex, and to measure its displacement vs. applied field (see Methods and Ref. \cite{Badea2015}).  This technique uses a 15-kHz AC magnetic field with peak amplitude of 1.25 mT to oscillate the vortex position by $\approx \pm 100$~nm, and measures the resulting change in magnetization. Here, this results in a positive signal $\Delta \theta_{\mathrm{K}}$ at the vortex core position, with an amplitude corresponding to the displacement due to the AC field. A static in-plane magnetic field can be applied with arbitrary direction using permanent magnets with automated motion control, and time-dependent fields are generated using the CPW. Figs.~\ref{setup}f-h show differential MOKE images of the vortex, as it is translated by increasing static field $B_{x}$.  The dark spot shows the location of the vortex core, which translates in the $-y$ direction as $B_{x}$ is increased. These measurements confirm that a single vortex domain is present and provide a measure of the vortex displacement susceptibility $\chi_0 \approx 80$~nm mT$^{-1}$. The sign of the displacement reveals the sign of the vortex helicity. By comparison to the PL scan in Fig.~\ref{setup}e, we can see that as the field is increased from $B_{x}=0$ to 9 mT, the vortex approaches close to the nanoparticle, then continues past.

\textbf{Probing the NV/vortex interaction.}  NV spins are initialized and read-out via the standard optically-detected magnetic resonance (ODMR) technique at room temperature\cite{Neumann2009,Jelezko2006}, using the same laser for NV excitation as is used for the MOKE measurements. A typical ODMR spectrum is produced by measuring the PL intensity while sweeping the MW frequency $f_{\mathrm{MW}}$.  Dips in the PL intensity reveal transitions between the $m_\mathrm{s} = 0$ to $m_\mathrm{s}= \pm1$ sublevels of the NV ground state.  For a single NV in a magnetic field, we generally expect to see two dips centered around the zero-field splitting $f_0=2.87$~GHz. See Methods for more detail on the techniques.

\begin{figure}[htbp]
	\centering
	\includegraphics[width=0.48\textwidth]{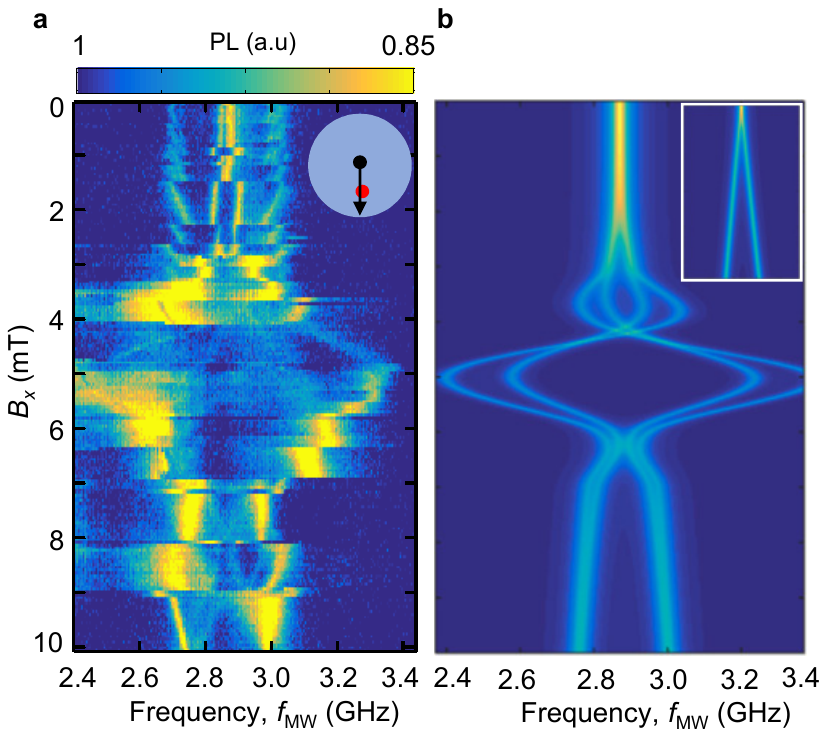}
	\caption{\textbf{Mapping the NV/vortex interaction. a}, Optically-detected magnetic resonance (ODMR) spectra vs. $B_x$ of the NVs contained within the DNP. As $B_x$ increases, the vortex (black dot) follows a path past the NVs (red dot) as illustrated in the inset. $\textbf{b}$, Simulated ODMR spectra of two NV centers with the same orientation, separated by 10~nm, 20~nm above the FM disk. Inset shows the same simulation but with no FM disk. The inset axes cover the same range as the main axes.}
	\label{Bsweeps}
\end{figure}

When the vortex core is brought near an NV (separation $d_{\mathrm{v-NV}}< 100$~nm), the NV spin experiences a significant magnetic field from the vortex.  Figure~\ref{Bsweeps}a shows ODMR vs. $f_{\mathrm{MW}}$ as the vortex is swept past the NVs by increasing the magnetic field $B_x$. The path of the vortex $\mathbf{r}_\mathrm{v}(B)$ is shown schematically in the inset (also see Fig.~\ref{setup}f-h). The largest splitting occurs around $B_x = 5$~mT, where the MOKE images show that $d_{\mathrm{v-NV}}$ is small. Here, the transitions to the $m_\mathrm{s}=+1$ and $m_\mathrm{s}=-1$ states are split by $\approx 900$~MHz, corresponding to a projection of $\approx 16$~mT along the NV symmetry axis. The orientations of the symmetry axes of these NVs are unknown, so this sets a lower bound on the actual magnetic field seen by the NV. Given the applied field $B_x=5$~mT, the vortex must be contributing $B\geq 11$~mT to the NV spin splitting. As the NV/vortex interaction increases, increased broadening of the resonance is also visible, with linewidths exceeding 100~MHz in some cases. We will show below that this can be explained by power-broadening due to strong vortex-induced amplification of the MW field. A distinct feature of the data in Fig.~\ref{Bsweeps}a is that the measured spin splittings do not change continuously with $B_x$, but instead undergo frequent steps. This is a result of vortex pinning caused by defects in the Permalloy, with steps occurring when the vortex becomes unpinned and jumps to a new equilibrium position\cite{Burgess2014}.  

The strength of the magnetic field gradient from the vortex can be roughly estimated from the slope $\gamma_{\mathrm{eff}}=\delta f/\delta B_x$ of the resonances vs. $B_x$ in Fig.~\ref{Bsweeps}a. (Specifically, $\gamma_{\mathrm{eff}}$ is related to the gradient in the direction of vortex motion.) Between $B_x=4$ and 5~mT, we observe the highest $\gamma_{\mathrm{eff}}\approx 320$~MHz mT$^{-1}$. For comparison, the splitting from the applied magnetic field alone is at most $\gamma = 28$~MHz mT$^{-1}$. The observed $\gamma_{\mathrm{eff}}$ corresponds to a gradient of the magnetic field projected on the NV axis $B' = \gamma_{\mathrm{eff}}/\gamma \chi_0 \approx 0.14$~mT nm$^{-1}$. For two NVs with the same axis, separated by 10s of nm, this gradient would cause splitting $\sim100$~MHz.

The general trends of the resonances vs. vortex position can be understood via comparison to simulation. We obtain the fringe field above a FM vortex domain using micromagnetic (OOMMF) simulation\cite{OOMMF}, as a function of applied field. We then calculate the NV resonances in that field, for NVs in particular locations, in particular orientations. Figure~\ref{Bsweeps}b shows the simulated resonances from a pair of NVs, both oriented with polar and azimuthal angles $(\phi,\theta)=\pi/4$. The two NVs are separated by 10~nm, and are located at positions $(x,y,z)=(10,-300,20)$~nm and $(20,-300,20)$~nm, relative to the center of the disk surface.  Note that this simulation does not take into account vortex pinning, or broadening of the resonances. The broadening may be due to coherent driving of spin-wave modes of the vortex, which is an effect not included in these static simulations. This will be further discussed below. Though the position and orientation of the measured NVs is not precisely known, the simulation captures some general features. The largest splitting is observed when $d_{\mathrm{v-NV}}$ is near its minimum, around $B_x=5$~mT.  In this region where the vortex is strongly coupled to the NVs, same resonance in different NVs are split $\sim100$~MHz.  When the vortex core is far from the NVs, the resonances tend towards the frequencies expected with no vortex present, shown for comparison in the inset. The experimental data show the same trends, though the experiment displays greater splitting when the vortex core is far from the NVs. This suggests that the fringe field contains higher gradients away from the core, perhaps due to irregularities in the FM film.

\begin{figure}[htbp]
	\centering
	\includegraphics[width=0.48\textwidth]{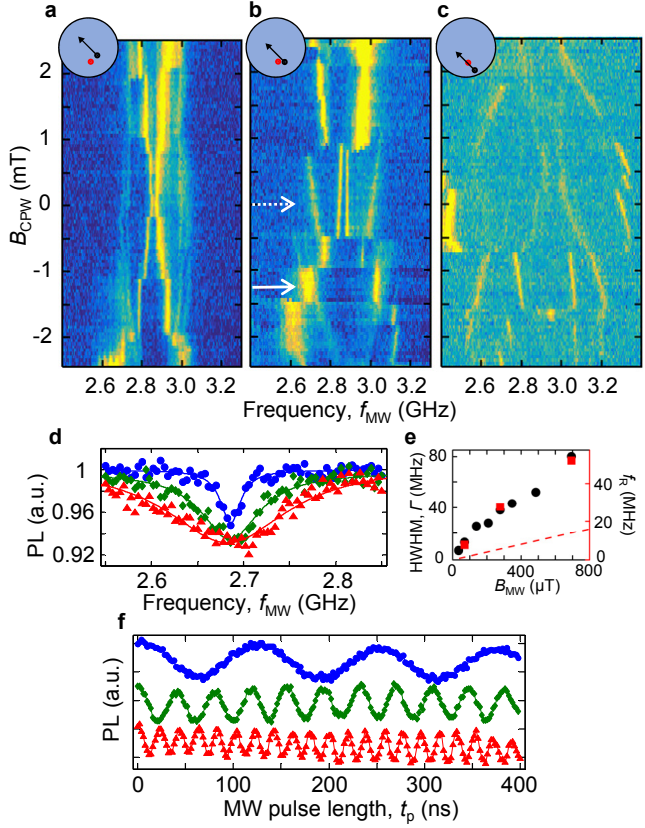}
	\caption{\textbf{Coherent control of vortex-coupled NV spins. a} - $\textbf{c}$, ODMR spectra with static $B_{x}$ plus variable $B_{\mathrm{CPW}}$ translating the vortex along the diagonal paths shown in the inset illustrations. $B_{x}=0,2 ,5$ mT respectively.  $\textbf{d}$, ODMR spectra with Lorentzian fits of the NV spin transition at the field indicated by the solid arrow in $\textbf{b}$ with $B_{\mathrm{MW}}= 70$~$\mu$T ($\bullet$), 280~$\mu$T ($\blacklozenge$) and 690~$\mu$T ($\blacktriangle$). $\textbf{e}$, (left axis, $\bullet$) Half-width at half-maximum $\mathit{\Gamma}$ of the resonance in \textbf{d} vs. $B_{\mathrm{MW}}$. (right axis, $\blacksquare$) Rabi frequencies $f_{\mathrm{R}}$ vs. $B_{\mathrm{MW}}$ from the data in $\textbf{f}$. Dashed line indicates upper bound of $f_{\mathrm{R}}$ with no vortex-induced enhancement. $\textbf{f}$, Rabi oscillations at the same resonance and the same three $B_{\mathrm{MW}}$ as in $\textbf{d}$.  The oscillations are offset for a clear comparison.} 
	\label{Bcpw}
\end{figure}

\textbf{Coherent manipulation of vortex-coupled NV spins.}  The first step towards fast addressability is to map the NV resonances in response to $B_{\mathrm{CPW}}$. Sweeping $B_{\mathrm{CPW}}$ with an additional static field $B_{x}$ moves the vortex along the paths illustrated in the insets to Fig.~\ref{Bcpw}a-c. The NV resonances that result from these vortex paths are shown by the corresponding ODMR scans (see Methods for more detail on the experimental protocol.) 

The ODMR scans in Fig.~\ref{Bcpw}a-c provide a more detailed look at the NV/vortex coupling. The resonances show a general trend of greater splitting at smaller $d_{\mathrm{v-NV}}$. The scan in c, closest to the vortex, was taken with reduced $B_{\mathrm{MW}}$, which reduces the power broadening that obscured the features at small $d_{\mathrm{v-NV}}$ in Fig.~\ref{Bsweeps}a. Sets of resonances with qualitatively different behavior vs. $B_{\mathrm{CPW}}$ likely arise from NVs with different orientation within the DNP. When lines are split into parallel doublets, this suggests that the transitions of NVs with the same orientation have become addressable.  From these maps, we can design a path of $\mathbf{r}_\mathrm{v}$ that will sequentially address one or more of the NV transitions.

To demonstrate that the broadening of the resonances is caused by vortex-enhanced power broadening, we measure both the linewidth $\mathit{\Gamma}$ (HWHM) and the Rabi frequency $f_\mathrm{R}$ of a spin transition at different $B_{\mathrm{MW}}$. Figure~\ref{Bcpw}d displays the ODMR spectrum of one of the transitions at the magnetic field indicated by the solid arrow in Fig.~\ref{Bcpw}b. The resonance dip is shown at increasing $B_{\mathrm{MW}}$ with a corresponding increase in linewidth.  A Lorentzian fit yields $\mathit{\Gamma}$ at these and other $B_{\mathrm{MW}}$, shown as circles in Fig.~\ref{Bcpw}e (left axis). $\mathit{\Gamma}$ is roughly linear with $B_{\mathrm{MW}}$, as expected if linewidth is dominated by MW-field-induced power broadening. We confirm the broadening mechanism by measuring coherent Rabi oscillations of this spin transition (see Methods). Figure~\ref{Bcpw}f shows Rabi oscillations with MW pulses at $f_{\mathrm{MW}}=2688$~MHz, and the same $B_{\mathrm{MW}}$ as used in (d). The Rabi frequency $f_\mathrm{R}$, extracted via a Fourier transform, is shown as squares in Fig.~\ref{Bcpw}e (right axis). For comparison, the line $f=\gamma B_{\mathrm{MW}}/\sqrt{2}$ ($\gamma = 28$~MHz mT$^{-1}$) is also shown, which is the maximum possible $f_\mathrm{R}$ for the NVs alone, in the absence of vortex-induced enhancement. Both $\mathit{\Gamma}$ and $f_\mathrm{R}$ show the same roughly linear trend vs. $B_{\mathrm{MW}}$. In the limit of strong MW-power broadening, the linewidth is expected to be $\mathit{\Gamma} = (\sqrt{\beta}/2)f_\mathrm{R}$, where $\beta$ is the ratio between optical excitation rate and spin initialization rate\cite{Dreau2011}. From the data in Fig.~\ref{Bcpw}e, we find $\beta \approx 10$, consistent with typical values for this ratio.  It is possible that this linewidth broadening and Rabi frequency enhancement are caused by coherent MW-induced excitation of spin wave modes in the disk, which in turn drive the spin transitions\cite{Park2005,VanderSar2015}.

\begin{figure}[htbp]
	\centering
	\includegraphics[width=0.47\textwidth]{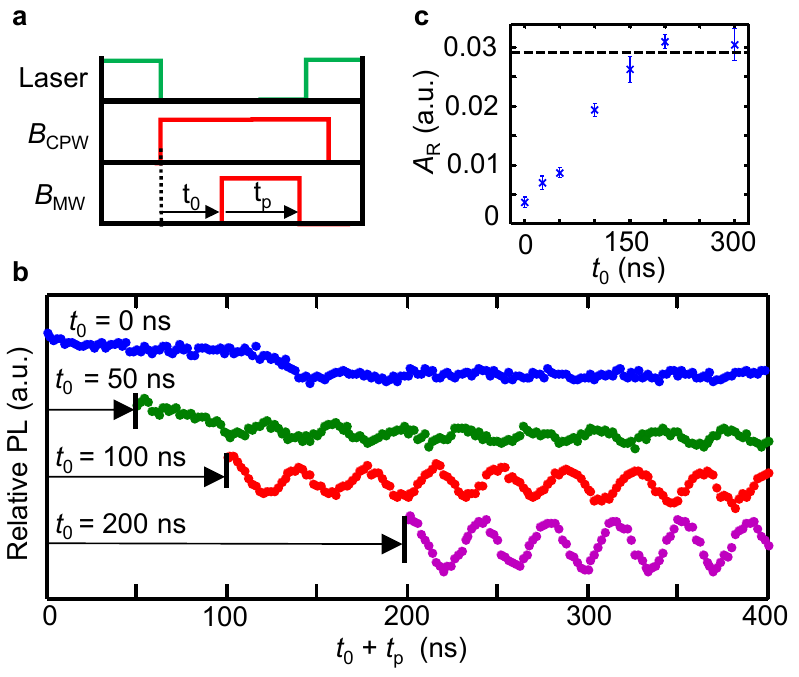}
	\caption{\textbf{Fast NV addressability and coherent control. a}, Pulse timing for NV initialization, and read-out (laser), vortex-enabled addressing ($B_{\mathrm{CPW}}$), and manipulation $B_{\mathrm{MW}}$. $t_0$ is the delay between $B_{\mathrm{CPW}}$  and $B_{\mathrm{MW}}$ and t$_{\mathrm{p}}$ is the $B_{\mathrm{MW}}$ pulse duration. $\textbf{b}$, Rabi oscillations following a vortex translation with different $t_0$. The oscillations are offset for a clear comparison. $\textbf{c}$, The Rabi amplitude, $A_\mathrm{R}$, vs. $t_0$. Dashed line indicates $A_\mathrm{R}$ with static vortex position. The error bars are determined by the sinusoidal fits.}
	\label{vortexjump}
\end{figure}

\textbf{Fast NV addressability and coherent manipulation of spins.}  We now demonstrate the capability to sequentially move the vortex and perform coherent operations on one of the NVs in the DNP.  The pulse sequence for this operation is shown in Fig.~\ref{vortexjump}a. The laser serves to first initialize, and finally measure all of the NV spins in the DNP.  During the spin initialization, $B_{\mathrm{CPW}}=0$, so the vortex is in the position indicated by the dashed arrow in Fig.~\ref{Bcpw}b. After the initialization pulse has ended, $B_{\mathrm{CPW}}$ is stepped to the solid arrow in Fig.~\ref{Bcpw}b. A MW pulse at $f_{\mathrm{MW}}=2688$~MHz then coherently rotates the spin. The MW pulse has duration $t_\mathrm{p}$, and is delayed from the $B_{\mathrm{CPW}}$ pulse by $t_0$.  From the map in Fig.~\ref{Bcpw}b, we see that at the initial vortex position, no NV transitions are resonant with $f_{\mathrm{MW}}$. After the step in $B_{\mathrm{CPW}}$ the vortex position undergoes some dynamics, finally relaxing to a position where a spin transition is now resonant with $f_{\mathrm{MW}}$.  The vortex position is reset after the spin read-out (resetting the vortex position before the read-out does not affect the outcome). 

The timescale for addressing an NV with the vortex and performing coherent spin rotation is demonstrated by the Rabi oscillations shown in Fig.~\ref{vortexjump}b. The ODMR contrast is plotted vs. $t_0+t_\mathrm{p}$ with different $t_{0}$.  Sinusoidal fits performed at $t_0+t_\mathrm{p}>200$~ns yield the oscillation amplitudes $A_\mathrm{R}$ (Fig.~\ref{vortexjump}c). When the MW pulse begins $t_0=200$~ns after the $B_{\mathrm{CPW}}$ step, clear Rabi oscillations are observed, with $A_\mathrm{R}$ equal to the value observed for static vortex position (dashed line in Fig.~\ref{vortexjump}c). As $t_0$ is reduced, the oscillation contrast is reduced. At $t_0 \leq 50$~ns, a non-oscillating decay of the signal is visible at $t_0 +t_\mathrm{p} \leq 100$~ns. The observed NV spin behavior can be understood in terms of the dynamic coupling of the vortex and NV transitions. The relaxation time for gyrotropic vortex dynamics in these structures is $\sim 50$~ns, as measured by time-resolved MOKE (not shown) and in agreement with previous work\cite{Chen2012}. Once the gyrotropic precession has fully decayed and the vortex has fully relaxed to its new equilibrium position at $t_0 \geq 200$~ns, the NV transition is on resonance, and coherent rotation proceeds as usual. While the vortex is still relaxing, the NV spin experiences a time-dependent spin splitting. If the MW pulse is present during these dynamics, the spin follows a more complex path on the Bloch sphere.

\section*{\textbf{Discussion}}
The NV/vortex system opens a path towards large-scale quantum registers, in which the vortex can be repositioned to address a single NV within the register. Specifically, tuning the vortex position to maximally split the spin transitions of one NV will separate those transitions from all other NVs in the register. As shown, the vortex can be repositioned on $\sim 100$~ns timescales, allowing for operations on many NVs within typical coherence times. Technical challenges that remain to be studied and addressed, however, include the effects of pinning on the freedom to position the vortex, and the possible presence of vortex-induced spin dephasing or decoherence.
	
It is interesting to ask whether the Bloch sphere dynamics that occur when the MW pulse is applied while the vortex is moving are coherent (as in the top curve of Fig.~\ref{vortexjump}b). The time evolution of the spin splitting here is similar to that required for adiabatic passage, suggesting that it may be possible to use the vortex dynamics to generate robust spin rotations. The coupled NV/vortex dynamics will provide a rich system for future study, opening new possibilities for coherent spin control with nanoscale addressability.

\section*{\textbf{Methods}}
\textbf{Experimental setup.}  ODMR and MOKE measurements were performed on a home-built scanning microscopy setup at room temperature.  A schematic of the setup is shown in Fig~\ref{setup}b. A CW 532 nm laser is sent through an acousto-optical modulator (AOM, Gooch and Housego) for laser pulsing and directed into an oil immersion objective (Olympus 100x 1.3 NA) by a dichroic mirror.  The objective focuses the laser onto the sample and both the PL and reflected laser are collected back through the same objective.  The reflected laser can be sent to the longitudinal MOKE measurement to map the magnetization component perpendicular to $B_{\mathrm{CPW}}$ ~\cite{Badea2015}. The magnetization maps are produced by raster scanning the disk while measuring the change in magnetization due to the 15-kHz applied $B_{\mathrm{CPW}}$ at each pixel. The PL proceeds to an electron-multiplication CCD (EMCCD) camera for imaging or sent to a pair of avalanche photodiodes (PDM-50ct) connected to a time-correlated single photon counting system (TCSPC, Hydraharp 400).  

\textbf{Sample preparation.} The sample is mounted on a 3D nanopositioning stage (Physik Instrumente P-517.3CL).  Permanent magnets are placed on automated translation stages that are positioned to move in the x- and y-directions indicated in Fig~\ref{setup}d and e. The static field is measured in situ by a 2D hall probe integrated in the sample holder.  The sample consists of a 200 nm thick gold coplanar waveguide (CPW) patterned via photo-lithography and thermal evaporation on a silicon substrate.  The CPW center conductor constricts to a $10 \mu$m width for $100 \mu$m at the middle.  In this region, Permalloy ($\text{Ni}_{0.81}\text{Fe}_{0.19}$) disks are fabricated via electron beam lithography, electron beam evaporation and liftoff, atop of the gold CPW. Lastly, diamond nanoparticles (DNPs) are dispersed over the entire surface.

The diamond nanoparticles (DiaScence, Quantum Particles), $\approx 25$ nm in diameter, are mixed with a 1.0$\%$ solution of polyvinyl alcohol, $M_{\mathrm{W}}$ 85,000-124,000 (Aldrich 363146) in de-ionized water.  Spin coating this solution directly on top of the CPW/Permalloy disks yielding a film thickness $\approx 35$ nm.

\textbf{Microwave field generation.} MW fields are produced by a signal generator (SRS SG386) and amplified (Pasternack PE15A4017). MW signals are either CW, internally modulated at low frequency, or pulsed at higher frequency using I/Q modulation.  An arbitrary function generator (Tektronix AFG3052C) and a digital delay pulse generator (SRS DG645) are used to provide signals at DC or with square amplitude modulation with fast rise time $\sim1$~ns. The MW field is combined with DC or square-modulated current with a resistive splitter (MicroCircuits ZFRSC-42-S+) and sent to the CPW.  The MW magnetic field (with amplitude referred to as $B_{\mathrm{MW}}$) produced by the CPW drives both the vortex and the NV spin transitions. The DC or low-frequency-modulated magnetic field (referred to as $B_{\mathrm{CPW}}$) shifts the NV transition frequencies, and moves the vortex.

\textbf{ODMR techniques.}  There are three types of ODMR scans performed in the main text. The first two methods involve simultaneous spin initialization, measurement, and incoherent MW-induced depolarization, while the third is a pulsed technique that permits coherent spin control. In all three, a timing scheme is used to provide signal and reference levels, as described below 

During an ODMR scan, the TCSPC continuously measures and time-tags each PL photon count in real time, $\{t_1,t_2,t_3...\}$.  The TCSPC also places event markers that are synced with the timing of the laser pulses and/or MW field modulation. A time interval is defined as the measurement time, and another interval is defined as the reference time (see below). The markers are then used to determine which counts arrived in which interval.

The generalized measurement described above is employed in three ways.
In the first method, NVs are under CW laser excitation with the static magnetic field fixed, while sweeping $f_{\mathrm{MW}}$, as in Fig.~\ref{Bsweeps}a. The MW field amplitude is modulated $B_{\mathrm{MW}}(t) = B_{\mathrm{MW}} M(t)$, where $M(t)$ is a square wave with one period
	
	\[ 
	M(t) =
	\begin{cases} 
	0 & 0\leq t < T_0/2 \\
	1 & T_0/2 \leq t < T_0  
	\end{cases}
	\]
	
At each $f_{\mathrm{MW}}$, many modulations occur (e.g. $T_0 = 1$~ms with measurement time = 1 s) and a marker is placed at the beginning of each period of $M$.  This allows us to separate the time-tagged PL counts, $\{t_i\}$, into bins such that the PL signal counts, $C_{\mathrm{sig}} = \sum\limits_{i}M(t_i)$, and PL reference counts, $C_{\mathrm{ref}} = \sum\limits_{i}(1-M(t_i))$. $C_{\mathrm{ref}}$ reflects the PL count rate when the spins are initialized, with no MW field applied. $C_{\mathrm{sig}}$ reflects the PL count rate when the spins are initialized, and also acted on by the MW field. The ODMR contrast is then defined as $S=C_{\mathrm{sig}}/C_{\mathrm{ref}}$. 
	
The second method is the same as the first, but we also modulate the vortex position via $B_{\mathrm{CPW}}$ (as in Fig.~\ref{Bcpw}a-c.) This is done to avoid thermal depinning of the vortex. When the vortex is at a position close to a transition between pinning configurations, it can stochastically jump to a new position, in a thermally activated process. Using the first method, the vortex may sit in such a position for many minutes. If a jump occurs during this time, discontinuities can appear in the data in a single $f_{\mathrm{MW}}$ scan. If the vortex can jump back and forth between two configurations, the scan features can be smeared. To avoid this, we want to reduce the dwell time of the vortex at these unstable positions. Modulating the vortex position back to some reference position essentially resets the vortex. It can still undergo thermal jumps, but if it does, the situation is quickly remedied. Here, both $B_{\mathrm{CPW}}(t) = B_{\mathrm{CPW}} M(t)$ and $B_{\mathrm{MW}}(t) = B_{\mathrm{MW}} M(t)$ are modulated at 1 kHz.  In this case, $C_{\mathrm{ref}}$ pertains to vortex position set by $B_{\mathrm{CPW}}=0$ with no MW applied, and $C_{\mathrm{sig}}$ pertains to the vortex with position set by the desired value of $B_{\mathrm{CPW}}$, with the MW field applied. Because the signal and reference here are measured in different static fields (from both $B_{\mathrm{CPW}}$ and the vortex field), $C_{\mathrm{ref}}$ may differ from $C_{\mathrm{sig}}$ even if no MW field is applied. To account for this, the ODMR contrast for each line scan is normalized to have the off-resonant contrast, $S=1$.
	
In the third method, pulsed measurements are carried out, separating the spin initialization, manipulation, and measurement in time. We use a measurement protocol with a period of $1.5~\mu$s.  The laser is on during the time interval $t_{\mathrm{ON}}=(0-1)~\mu$s and off during the interval $t_{\mathrm{OFF}}=(1-1.5)~\mu$s. The measurement time interval (read-out) is defined as $t_{\mathrm{m}}=(0-300)$ ns and the reference time interval is defined as $t_{\mathrm{r}}=(700-1000)$ ns.  $C_{\mathrm{sig}}$ is then the number of $\{ t_i \}$ in the range $t_{\mathrm{m}}$ and $C_{\mathrm{ref}}$ is the number of $\{t_i\}$ in the range $t_{\mathrm{r}}$. The MW field is also pulsed and is on either during $t_{\mathrm{OFF}}$ (for coherent manipulation, as in Fig.~\ref{Bcpw}f and \ref{vortexjump}b) or during $t_{\mathrm{ON}}$ (for incoherent depolarization, as in Fig.~\ref{Bcpw}d). 

\acknowledgements
This work was supported by the U.S. Department of Energy, Office of Science, Basic Energy Sciences, under Award \#DE-SC008148. 

\section*{Author Contribution}
J.B. conceived and supervised the research, analyzed data and performed simulations. R.B. designed and fabricated the samples and performed the MOKE measurements. M.S.W. set up and performed the NV measurements. J.B. and M.S.W. wrote the manuscript and designed the figures. 

\section*{Competing financial interests}
The authors declare no competing financial interests.

\end{document}